\documentclass[11pt]{article}
\usepackage[T1]{fontenc}
\usepackage[utf8]{inputenc}
\usepackage{amsmath,amssymb,amsthm}
\usepackage[margin=1in]{geometry}
\usepackage{fancyvrb}
\usepackage[hidelinks]{hyperref}
\numberwithin{equation}{section}
\newtheorem{theorem}{Theorem}[section]

\theoremstyle{remark}\newtheorem{remark}[theorem]{Remark}

\title{Existence of a third self-similar  solution\\ of the Yang--Mills equation
in high even dimensions}
\author{%
Piotr Bizo\'n\thanks{Email: \texttt{piotr.bizon@uj.edu.pl}}\\[1ex]
 Institute of Theoretical Physics,
Jagiellonian University, Krak\'ow
}
\date{}
\begin{document}\maketitle
\begin{abstract}
We prove that the spherically symmetric $SO(d)$ Yang--Mills equation admits a third
smooth self-similar profile, in addition to the two
known explicit profiles, in every even dimension
$d\geq 12$. This complements the analogous existence theorem
proved in~\cite{BK} in every odd dimension $d\geq 11$.
In contrast to the odd-dimensional case, where the
classification of smooth self-similar profiles reduces
to counting roots of an explicit resonance
polynomial~\cite{BGW}, the even-dimensional proof requires
solving a nonlinear connection problem. In $d=10$, where one explicit profile degenerates
to the zero solution, we show by formal methods
the existence of a second nontrivial smooth
profile in addition to the single nontrivial explicit one.
\end{abstract}
\section{Introduction}\label{sec:picture}
 Self-similar solutions of the Yang–Mills equation provide natural models of finite-time singularity formation. For such solutions a fixed spatial profile is compressed into a shrinking region around the blowup point. Constructing and classifying these profiles is  a first step towards understanding the possible local blowup dynamics. Their role in the evolution depends  on their  stability.

Under the spherically symmetric magnetic ansatz with gauge group $\mathrm{SO}(d)$, the Yang--Mills equation in $(d+1)$-dimensional Minkowski spacetime reduces to the radial nonlinear wave equation
\begin{equation}\label{eq:YMwave}
\phi_{tt}=\phi_{rr}+\frac{d-3}{r}\phi_r
+\frac{d-2}{r^2}\phi(1-\phi^2).
\end{equation}
The equation is invariant under $\phi(t,r)\mapsto\phi(t/\lambda,r/\lambda)$ for any $\lambda>0$. Solutions unchanged by this scaling are called self-similar. By time-translation invariance, for any $T$ they take the form
\[
\phi(t,r)=u(y),\qquad y=\frac{r}{T-t},\qquad t<T.
\]
Substituting this ansatz into~\eqref{eq:YMwave} gives the profile equation
\begin{equation}\label{eq:profile}
y^2(1-y^2)u''+\bigl((d-3)y-2y^3\bigr)u'
+(d-2)u(1-u^2)=0.
\end{equation}
The profiles of interest are nonconstant and smooth on $0\le y\le1$,
so the associated solutions are smooth in the causal past of $(T,0)$,
where they blow up as their spatial scale $T-t$ shrinks to zero.
 Throughout the paper, profiles related by the reflection $u\mapsto-u$ are regarded as equivalent.
\vskip 0.1cm
A standard way to solve this two-endpoint boundary-value problem is shooting. Near the origin, there is a one-parameter family of regular local solutions
\begin{equation}\label{eq:origin}
u(y;a)=1-ay^2+O(y^4),\qquad a\in\mathbb R.
\end{equation}
The shooting method consists in continuing these solutions towards
$y=1$ and selecting the values of the parameter $a$ for which the resulting shot
is smooth at the light cone $y=1$.

As shown in~\cite{BGW}, this shooting problem simplifies
considerably when $d\geq 10$. More precisely,  for each fixed $d$ in this range and every $a>0$, the corresponding solution $u(y;a)$ extends to the whole interval $0<y<1$, where it is positive, strictly decreasing, and has a finite limit
\[
c(a):=\lim_{y\to1^{-}}u(y;a).
\]
Moreover, the shooting map $a\mapsto c(a)$ is a strictly decreasing bijection from $[0,\infty)$ onto $(0,1]$, with $c(0)=1$ and 
$
\lim_{a\to\infty}c(a)=0.
$
Thus, the oscillations and repeated endpoint values that complicate shooting in lower dimensions do not occur here: each endpoint value in $(0,1]$ corresponds to exactly one shooting parameter. Below, we take these results as input. It remains to determine the values of the shooting parameter $a$ for which the solution $u(y;a)$ extends smoothly to $y=1$. 

Two such solutions are known explicitly:
\begin{equation}\label{explicit}
u_\pm(y)=1-\frac{\alpha_\pm y^2}{y^2+\beta_\pm},\qquad
 a_\pm=\frac{\alpha_\pm}{\beta_\pm},\qquad 0<a_+<a_-,
\end{equation}
where, using $r=\sqrt{3(d-4)(d-2)}$,
\[
\alpha_\pm=2\pm\frac{2r}{3(d-2)},\qquad \beta_\pm=\frac{2(d-4)\pm r}{3}.
\]
The solutions $u_{+}$ and $u_{-}$ exist for $d\geq5$ and $d\geq11$, respectively.

 The question is whether there are any other smooth profiles.
 The analysis depends crucially on the parity of $d$
because the local behavior of solutions near $y=1$
is governed by the indicial exponents $0$ and

\[
\rho=\frac{d-3}{2}.
\]
 The problem is therefore resonant
when $d$ is odd, in which case $\rho$ is an integer,
and nonresonant when $d$ is even, in which case $\rho$
is a half-integer. We discuss these two cases separately.

\paragraph{Odd dimensions}

For odd $d=2\ell +5\geq  11$ the general local solution
with finite endpoint value $c$ has the expansion
\[
u(y)
=
c+\sum_{n=1}^{\infty}\alpha_n (1-y)^n
+A_\ell (c) (1-y)^{\ell +1} \log(1-y)
+\cdots,
\]
where we omitted higher-order power--log terms
$(1-y)^j \log^k(1-y)$.
Here $c$ and $\alpha_{\ell +1}$ are free parameters, while
$\alpha_1,\ldots,\alpha_\ell $ are determined recursively by $c$.
The coefficient
\[
A_\ell (c)=c(1-c^2)P_\ell (c^2),
\]
measures the obstruction to continuing
the Taylor expansion through the resonant order $(1-y)^{\ell +1}$. Here $P_\ell $ is an explicitly computable polynomial of degree $\ell $,
defined up to a constant factor.
If $P_\ell (c^2)=0$, all logarithmic terms vanish and the expansion
reduces to a convergent Taylor series. Since the origin-regular shots have $0<c<1$, they are smooth at $y=1$ if and only if $P_\ell (c^2)=0$. Combined with monotonicity of the shooting map, this reduces the classification of smooth self-similar profiles to counting distinct roots of $P_\ell (c^2)$ in $0<c^2<1$: each root determines one smooth profile, modulo reflection.
Two roots correspond to the explicit profiles.
The existence of at least one additional root was
recently proved in~\cite{BK}, establishing the existence
of a third nonconstant smooth profile in every odd
dimension $d\geq 11$ and confirming the conjecture
formulated in~\cite{BGW}.

\paragraph{Even dimensions.}
Write $d=2m$ with an integer $m\ge5$. Since
$\rho=m-\frac32$ is not an integer, each endpoint value $c$
determines a unique local solution $u_0(y;c)$ analytic at $y=1$.
The general local solution with the same endpoint value has an
additional free parameter, the amplitude of its leading
nonanalytic term. This contrasts with the logarithmic term in
odd dimensions, whose coefficient is determined by $c$.

For an origin-regular shot, continuation from the origin to the
light cone selects both the endpoint value $c(a)$ and the
nonanalytic amplitude $B_m(a)$. The resulting endpoint expansion
takes the form
\begin{equation}
u(y;a)=u_{\mathrm{an}}(y;a)
+B_m(a)(1-y)^\rho g(y;a),
\label{eq:endpoint-expansion}
\end{equation}
where $u_{\mathrm{an}}$ and $g$ are analytic at $y=1$, with
$u_{\mathrm{an}}(1;a)=c(a)$ and $g(1;a)=1$.
Here $u_{\mathrm{an}}$ denotes the analytic component of the
full solution, not in general the analytic solution
$u_0(y;c(a))$. The decomposition and its dependence on the
shooting parameter are justified in \ref{app:endpoint}.
By~\eqref{eq:endpoint-expansion}, the shot is
smooth at the light cone if and only if
$B_m(a)=0$.
Unlike in odd dimensions, this criterion
cannot be reduced to an algebraic condition
on the endpoint value $c$.
Instead, finding the zeros of $B_m$ requires
controlling the nonanalytic amplitude along
the family of origin-regular shots.
This is a genuinely nonlinear connection problem.

For $m\geq6$, establishing a third smooth
profile amounts to finding a zero of $B_m$
distinct from the two explicit zeros,
\[
B_m(a_+)=B_m(a_-)=0.
\]
For $m=5$, only the explicit zero $a_+$
remains, and we seek a second nontrivial
smooth profile.
By continuity of $B_m$, it is enough to find
a sign change in an interval containing
no explicit zero.
The argument splits into three cases:
\begin{itemize}
\item For $m=6,7,8$, $B_m$ crosses both explicit
zeros in the same direction, forcing an additional
zero in $(a_+,a_-)$.

\item For every integer $m\geq9$, the sign of $B_m$
immediately beyond $a_-$ differs from its sign for
large $a$, forcing an additional zero in
$(a_-,\infty)$.

\item For $m=5$, the sign of $B_5$ immediately
beyond $a_+$ differs from its sign for large $a$,
forcing an additional zero in $(a_+,\infty)$.
\end{itemize}
The first two cases are proved rigorously, while
the third is established only at a formal level.
Thus, we prove the existence of a third nontrivial
smooth profile in every even dimension $d=2m\geq12$.
For $m=5$, where one explicit profile degenerates
to the zero solution, we show by formal methods
the existence of a second nontrivial smooth
profile in addition to the remaining explicit one.

The rest of the paper is organized as follows.
In Section~\ref{sec:large}, we determine the
large-$a$ behavior of $B_m(a)$, including the
exceptional case $m=5$.
In Section~\ref{sec:signs}, we determine the
crossing directions at the explicit zeros
and complete the existence proof for $m\geq6$.
The technical details are collected
in the appendices.

\section{Large-$a$ behavior}
\label{sec:large}

\subsection{The case $m\geq6$}
\label{sec:large-high}
We use the shorthands
\begin{equation}\label{eq:exponents}
\nu=\sqrt{m^2-6m+6},
\qquad
k=m-2-\nu,
\end{equation}
so that $1<k<2$. As before, $\rho=m-\frac32$ is the  half-integer 
 light-cone indicial exponent.

The origin expansion~\eqref{eq:origin} identifies
$y\sim a^{-1/2}$ as the scale on which the deviation
from $u=1$ is no longer small. To resolve this
shrinking region, let
\[
x=\sqrt a\,y,
\qquad
U_a(x)=u(y;a).
\]
The rescaled profile satisfies
\[
x^2\left(1-\frac{x^2}{a}\right)U_a''
+\left((2m-3)x-\frac{2x^3}{a}\right)U_a'
+(2m-2)U_a(1-U_a^2)=0,
\qquad
U_a(x)=1-x^2+O(x^4).
\]
For fixed $x$, letting $a\to\infty$ gives
\begin{equation}\label{eq:static}
x^2Q''+(2m-3)xQ'+(2m-2)Q(1-Q^2)=0,
\qquad Q(x)=1-x^2+O(x^4).
\end{equation}
The normalized solution $Q$ is the limiting core profile:
$U_a\to Q$ on compact $x$-intervals as $a\to\infty$.
With $\tau=\log x$ and $q(\tau)=Q(e^\tau)$,
equation~\eqref{eq:static} becomes
\[
\ddot q+(2m-4)\dot q+(2m-2)q(1-q^2)=0.
\]
The origin expansion shows that the orbit enters the
forward-invariant strip
\[
0<q<1,\qquad -kq<\dot q<0,
\]
whose invariance follows by checking the vector field
on its boundary. Thus $Q$ is global, positive, and
strictly decreasing to zero. Linearization at $q=0$ gives
\begin{equation}\label{eq:static-tail}
Q(x)\sim C_m x^{-k}\qquad\text{as }x\to\infty,
\qquad C_m>0.
\end{equation}
Indeed, the inequality $\dot q+kq>0$ makes
$e^{k\tau}q(\tau)$ strictly increasing and hence rules
out decay along the faster mode.

The outer field inherits its amplitude and small-$y$
behavior from this tail in the overlap region
$a^{-1/2}\ll y\ll1$:
\[
u(y;a)=U_a(\sqrt a\,y)
\simeq Q(\sqrt a\,y)
\sim C_m a^{-k/2}y^{-k}.
\]
To extend this approximation beyond the overlap
region, we substitute the ansatz
\begin{equation}\label{eq:outerapprox}
u(y;a)\simeq C_m a^{-k/2}F_m(y)
\end{equation}
into~\eqref{eq:profile} and retain the leading terms
as $a\to\infty$. This gives the linear equation
\begin{equation}\label{eq:outer-linear}
y^2(1-y^2)F_m''
+\bigl((2m-3)y-2y^3\bigr)F_m'
+(2m-2)F_m=0.
\end{equation}
Matching the behavior $F_m(y)\sim y^{-k}$ for $y\to0$ selects the solution
\begin{equation}\label{eq:F}
F_m(y)=y^{-k}{}_2F_1\!\left(
-\frac{k}{2},\frac{1-k}{2};
1+\nu;y^2
\right).
\end{equation}
Thus the core fixes the amplitude, while the linear
outer equation fixes the shape.

Applying the hypergeometric connection formula
\cite[Eq.~(15.8.4)]{DLMF} to~\eqref{eq:F} gives
\[
F_m(y)=F_{m,\mathrm{an}}(y)
+
\mathcal K_m
\,y^{-k}(1-y^2)^\rho
{}_2F_1\!\left(
\frac{m+\nu}{2},\frac{m-1+\nu}{2};
1+\rho;1-y^2
\right),
\]
where $F_{m,\mathrm{an}}$ is analytic at $y=1$ and
\begin{equation}\label{eq:K}
\mathcal K_m=
\frac{\Gamma(1+\nu)\Gamma(-\rho)}
     {\Gamma(-k/2)\Gamma((1-k)/2)}.
\end{equation}
Expanding at $y=1$, we obtain
\begin{equation}\label{eq:Fconnection}
F_m(y)=F_{m,\mathrm{an}}(y)
+2^{\rho} \mathcal K_m(1-y)^\rho\bigl(1+O(1-y)\bigr).
\end{equation}
Comparing this expansion with~\eqref{eq:endpoint-expansion}
and recalling the outer amplitude $C_ma^{-k/2}$, we obtain the asymptotic behaviour for $a\to\infty$
\begin{equation}\label{eq:B-asymptotic}
B_m(a)\sim 2^{\rho} C_m\mathcal K_ma^{-k/2}.
\end{equation}
The matching and the passage to the light-cone
coefficient are justified rigorously in
\ref{app:matching}.

It remains to determine the sign of $\mathcal K_m$.
Since $1<k<2$, both denominator Gamma factors
in~\eqref{eq:K} are negative. Hence
\begin{equation}\label{eq:eventual-sign}
\operatorname{sgn}\mathcal K_m
=\operatorname{sgn}\Gamma(-\rho)
=\varepsilon_m,
\qquad
\varepsilon_m:=(-1)^{m-1}.
\end{equation}
Since $C_m$ is positive, this is also the sign of $B_m(a)$
for sufficiently large $a$: negative when
$m=6,8,10,\ldots$, and positive when
$m=7,9,11,\ldots$.

\subsection{The exceptional dimension $d=10$ ($m=5$)}
\label{sec:d10}

For $m=5$, the limiting core profile is explicit:
\begin{equation}\label{eq:Q10}
Q(x)=\frac{1}{1+x^2}.
\end{equation}
In the overlap region $a^{-1/2}\ll y\ll1$,
\[
Q(\sqrt a\,y)
=a^{-1}y^{-2}-a^{-2}y^{-4}
+O(a^{-3}y^{-6}).
\]
The leading outer term is therefore of order
$a^{-1}$. Since the cubic term
in~\eqref{eq:profile} first contributes at order
$a^{-3}$, the first two outer terms satisfy
the linear equation
\[
y^2(1-y^2)F''+(7y-2y^3)F'+8F=0.
\]
One solution is
\[
F_5(y)=y^{-2}+\frac14.
\]
Reduction of order gives a second linearly
independent solution,
\begin{equation}\label{eq:G10}
G(y)=-2F_5(y)\int_1^y
\frac{s^{-7}(1-s^2)^{5/2}}{F_5(s)^2}\,ds.
\end{equation}
The factor $-2$ normalizes the leading term
at the origin to $y^{-4}$, while the lower
integration limit fixes the freedom to add
a multiple of $F_5$ by imposing $G(1)=0$.
As $y\to0$,
\[
G(y)=y^{-4}+6y^{-2}\log y+O(y^{-2}).
\]
Matching the leading core term selects
$a^{-1}F_5$, while the term $-a^{-2}y^{-4}$
fixes the coefficient of $G$ to be $-a^{-2}$.
The accompanying logarithmic term is matched
by the first inner correction and introduces
a logarithmic dependence on $a$ in the coefficient
of $F_5$. Since $F_5$ is analytic at $y=1$,
this does not affect $B_5(a)$.

Using $F_5(y)=5/4+O(1-y)$ and
$1-s^2=2(1-s)\bigl(1+O(1-s)\bigr)$ in the
integral representation of $G$, we obtain
\[
G(y)=\frac{32\sqrt2}{5}
\int_y^1(1-s)^{5/2}\,ds\,
\bigl(1+O(1-y)\bigr).
\]
Hence
\begin{equation}\label{eq:G10endpoint}
G(y)\sim \frac{64\sqrt2}{35}(1-y)^{7/2}\qquad\mbox{as}\quad
 y\to1^-.
\end{equation}
and therefore 
\begin{equation}\label{eq:B10}
B_5(a)\sim-\frac{64\sqrt2}{35}\,a^{-2}
\qquad\mbox{as}\quad a\to\infty.
\end{equation}
\begin{remark}
A rigorous proof of~\eqref{eq:B10} requires
controlling the first inner correction and
propagating the matching error to the outer
region. These standard but lengthy estimates
are omitted to avoid obscuring the main argument.
We therefore state \eqref{eq:B10} as a formal
asymptotic prediction.
\end{remark}

\section{Additional profiles from crossing directions}
\label{sec:signs}

The direction in which $B_m$ crosses an explicit
zero is determined by the variation
\[
w(y):=\left.
\frac{\partial u(y;a)}{\partial a}
\right|_{a=a_\pm},
\]
so that
\[
u(y;a_\pm+\delta a)
=u_\pm(y)+\delta a\,w(y)+O(\delta a^2).
\]
Differentiating~\eqref{eq:profile} with respect
to $a$ at $a=a_\pm$ gives
\begin{equation}\label{variation}
y^2(1-y^2)w''
+\bigl((2m-3)y-2y^3\bigr)w'
+(2m-2)(1-3u_\pm^2)w=0.
\end{equation}
The origin expansion $u(y;a)=1-ay^2+O(y^4)$
gives $w(y)=-y^2+O(y^4)$.
Similarly, differentiating the endpoint
expansion~\eqref{eq:endpoint-expansion}
at $a=a_\pm$ and using $B_m(a_\pm)=0$ gives
\begin{equation}\label{eq:w-endpoint}
w(y)=w_{\rm an}(y)
+B_m'(a_\pm)(1-y)^\rho g(y;a_\pm),
\end{equation}
where $w_{\rm an}$ is analytic at $y=1$.
Since $g(1;a_\pm)=1$, the leading fractional
coefficient of $w$ is $B_m'(a_\pm)$.
This coefficient determines the crossing
direction whenever it is nonzero, because
\[
B_m(a_\pm+\delta a)
=\delta a\,B_m'(a_\pm)+O(\delta a^2).
\]
We carry out this analysis for both explicit
profiles when $m\geq6$ and for the single
explicit profile $u_+$ when $m=5$.

For either branch, suppress the sign label
and set
\begin{equation}\label{transform}
A=1+\beta,\qquad
\xi=\frac{Ay^2}{y^2+\beta},\qquad
w=-\frac{\beta}{A}\xi f(\xi).
\end{equation}
This maps $0\leq y\leq1$ to $0\leq\xi\leq1$,
and the origin condition $w=-y^2+O(y^4)$ becomes
$f(0)=1$.
Inserting~\eqref{transform} into~\eqref{variation}
and using the explicit profile parameters gives
\begin{equation}\label{heun}
f''
+\left(
\frac{m+1}{\xi}
-\frac{m-\frac52}{\xi-1}
+\frac{1}{2(\xi-A)}
\right)f'
+\frac{m+\frac{3A-1}{2}-4\xi}
{\xi(\xi-1)(\xi-A)}f=0.
\end{equation}
Equation~\eqref{heun} is a Heun equation with
four regular singular points at
$\xi=0,1,A,\infty$.

The origin-regular solution has the Taylor expansion
\begin{equation}\label{eq:f-taylor}
f(\xi)=\sum_{n=0}^{\infty}f_n\xi^n,
\qquad f_0=1.
\end{equation}
Since $A>1$, the nearest possible singularity
is at $\xi=1$, so the series converges
for $0\leq \xi<1$.

Near the light cone, the origin-regular solution
defined by~\eqref{eq:f-taylor} has the representation
\begin{equation}\label{eq:f-endpoint}
f(\xi)=f_{\rm an}(\xi)
+\mathcal C(1-\xi)^\rho h(\xi),
\qquad h(1)=1,
\end{equation}
where $f_{\rm an}$ and $h$ are analytic at $\xi=1$.
Comparing this with~\eqref{eq:w-endpoint}
using the transformation~\eqref{transform},
we obtain, after restoring the branch labels,
\begin{equation}\label{slope}
B_m'(a_\pm)
=-\frac{\beta_\pm}{A_\pm}
\left(\frac{2\beta_\pm}{A_\pm}\right)^\rho
\mathcal C_{m,\pm}.
\end{equation}

Our task is to determine the sign of the
connection coefficient $\mathcal C$.
To this end, we differentiate~\eqref{eq:f-endpoint}
$m-1$ times with respect to $\xi$.
Recalling that $\rho=m-\frac32$, we obtain
\begin{equation}\label{eq:f-endpoint-derivative}
f^{(m-1)}(\xi)
=\varepsilon_m
\left(\prod_{j=0}^{m-2}(\rho-j)\right)\,
\mathcal C(1-\xi)^{-1/2}+O(1),
\qquad
\varepsilon_m:=(-1)^{m-1},
\end{equation}
where the product is positive. Thus, if $\mathcal C\ne0$,
the derivative diverges with sign
$\varepsilon_m\operatorname{sgn}\mathcal C$.

On the other hand, differentiating the Taylor
series~\eqref{eq:f-taylor} $m-1$ times gives
\begin{equation}\label{eq:f-taylor-derivative}
f^{(m-1)}(\xi)
=\sum_{n\geq m-1}
\frac{n!}{(n-m+1)!}\,f_n\xi^{n-m+1}.
\end{equation}
The analysis of the recurrence relation in
\ref{app:coefficients} shows that
all sufficiently large Taylor coefficients
$f_n$ have the same sign
$\sigma\in\{-1,1\}$ and that $f^{(m-1)}$
diverges as $\xi\to1^-$.
By~\eqref{eq:f-taylor-derivative}, this
divergence has sign $\sigma$, since the
finitely many terms preceding the tail
remain bounded.
Comparing with~\eqref{eq:f-endpoint-derivative},
we conclude that $\mathcal C\ne0$ and
\begin{equation}\label{sigma}
\operatorname{sgn}\mathcal C=\sigma \varepsilon_m.
\end{equation}
Since the prefactor in~\eqref{slope} is negative,
we obtain
\begin{equation}\label{eq:coefficient-sign}
\operatorname{sgn}B_m'(a_\pm)
=-\sigma\varepsilon_m.
\end{equation}
The appendix establishes $\sigma=-1$ on the
plus branch for every $m\geq5$ and on the
minus branch for $m=6,7,8$, whereas $\sigma=1$
on the minus branch for $m\geq9$. The resulting crossing signs are summarized
below, together with the intervals containing
an additional zero.
For $m\geq9$, the conclusion also uses
\eqref{eq:B-asymptotic} and
\eqref{eq:eventual-sign}; for $m=5$, it relies
on the formal prediction~\eqref{eq:B10}.

\begin{equation}\label{signplan}
\begin{array}{c|c|c|c}
m&
\operatorname{sgn}B_m'(a_+)&
\operatorname{sgn}B_m'(a_-)&
\text{interval containing a new zero}\\ \hline
6,7,8&
\varepsilon_m&\varepsilon_m&(a_+,a_-)\\
m\geq9&
\varepsilon_m&-\varepsilon_m&(a_-,\infty)\\
5&+1&\text{---}&(a_+,\infty)
\end{array}
\end{equation}
For $m\geq6$, these sign changes and
continuity of $B_m(a)$ yield the following
existence result.

\begin{theorem}\label{conclusion}
For every integer $m\geq6$ (spatial dimension
$d=2m$), there are at least three
nonconstant smooth self-similar profiles
on $[0,1]$, modulo reflection.
Besides $u_\pm$, there is a profile $u_\star$
with shooting parameter
\[
a_\star\in(a_+,a_-)\quad(m=6,7,8),
\qquad
a_\star\in(a_-,\infty)\quad(m\geq9).
\]
\end{theorem}

\begin{proof}
For $m=6,7,8$, the derivatives of $B_m$
at the two explicit zeros have the same
nonzero sign,
hence $B_m(a_++\delta)$ and $B_m(a_--\delta)$
have opposite signs for sufficiently small
$\delta>0$.
Continuity therefore gives an additional zero
in $(a_+,a_-)$.

For $m\geq9$, we have
$\operatorname{sgn}B_m'(a_-)=-\varepsilon_m$,
so $B_m(a_-+\delta)$ has sign
$-\varepsilon_m$ for sufficiently small
$\delta>0$.
On the other hand,
\eqref{eq:B-asymptotic} and
\eqref{eq:eventual-sign} give sign
$\varepsilon_m$ for $B_m(a)$ at a sufficiently
large parameter.
Continuity therefore gives an additional zero
in $(a_-,\infty)$.

In both cases, Theorem~\ref{thm:endpoint-dependence}
implies smoothness.
\end{proof}

For $m=5$, the crossing sign $B_5'(a_+)>0$
gives $B_5(a_++\delta)>0$ for sufficiently
small $\delta>0$, whereas~\eqref{eq:B10}
predicts that $B_5(a)<0$ for large $a$.
Taking this prediction as input, continuity
forces an additional zero in $(a_+,\infty)$.
Theorem~\ref{thm:endpoint-dependence} then gives
a second nontrivial smooth profile.
This conclusion remains formal because the
error estimate needed to justify~\eqref{eq:B10}
has not been established.

\appendix
\renewcommand{\thesection}{Appendix~\Alph{section}}
\renewcommand{\thesubsection}{\Alph{section}.\arabic{subsection}}
\renewcommand{\theequation}{\Alph{section}.\arabic{equation}}
\renewcommand{\thetheorem}{\Alph{section}.\arabic{theorem}}
\section{Endpoint coordinates}
\label{app:endpoint}

We describe a shot near the light cone by two numbers: its limiting
value $c=u(1)$ and the amplitude $B$ of the leading fractional-power
term. Smooth profiles are precisely those for which $B=0$.
We construct these endpoint coordinates and show that they depend
analytically on the shooting parameter.

\subsection{Local behaviour at the light cone}

For a fixed integer $m\ge5$, put
\[
s=\frac1y-1,\qquad
v(s)=u(y),\qquad
\rho=m-\frac32.
\]
The light cone corresponds to $s=0$, and the profile equation becomes
\begin{equation}
s(2+s)v''-2(\rho-1)(1+s)v'
+(2m-2)v(1-v^2)=0.
\label{eq:endpoint-ode}
\end{equation}
The indicial exponents are $0$ and $\rho$.

We use the following theorem of
Breitenlohner--Forg\'acs--Maison (BFM)~\cite{BFM}.
\begin{theorem}[BFM]
\label{thm:BFM}
The system
\[
sX'=sf(s,X,Y),\qquad
sY'=-\Lambda Y+sg(s,X,Y),
\]
where $f$ and $g$ are analytic and $\Lambda$ is a positive constant,
has a local solution near $s=0$ satisfying
\[
X(0)=c,\qquad Y(0)=0.
\]
The solution is analytic in $s$, $c$, and any additional parameters
entering the equations analytically, and is unique among solutions
continuous at $s=0$ with these limiting values.
\end{theorem}

We apply this theorem twice: first to construct the analytic
solution with prescribed endpoint value $c$, and then to include
the independent fractional-power correction.

\paragraph{The analytic solution.}
We first construct the solution analytic at $s=0$ with $v(0)=c$.
To bring equation~\eqref{eq:endpoint-ode}  into BFM form, we seek a polynomial
\[
P(s,v)=\sum_{j=0}^{m-2}s^jp_j(v)
\]
which approximates the relation between $v'$ and $v$ along
analytic solutions.
Substituting
\[
v'=P(s,v),\qquad v''=P_s+P_vP
\]
into~\eqref{eq:endpoint-ode} leaves the residual
\[
\mathcal R[P](s,v)
:=
s(2+s)(P_s+P_vP)
-2(\rho-1)(1+s)P
+(2m-2)v(1-v^2).
\]
This measures the failure of the relation $v'=P(s,v)$ to satisfy
the profile equation. We choose the coefficients $p_j(v)$ to
cancel successive powers of $s$, so that
$\mathcal R[P]=O(s^{m-1})$.

Cancellation of the constant term gives
\[
p_0(v)=\frac{2m-2}{2m-5}v(1-v^2),
\]
while cancellation of the coefficient of $s$ gives
\[
2(2-\rho)p_1+2p_0p_0'-2(\rho-1)p_0=0,
\]
which determines $p_1$ from $p_0$.
More generally, suppose that $p_0,\ldots,p_{j-1}$ have been
chosen so that
\[
P_{j-1}(s,v)=\sum_{k=0}^{j-1}s^kp_k(v),
\qquad
\mathcal R[P_{j-1}]=s^jr_j(v)+O(s^{j+1}).
\]
Then
\[
\mathcal R[P_{j-1}+s^jp_j]
=s^j\bigl[r_j+2(j+1-\rho)p_j\bigr]+O(s^{j+1}),
\]
so the next coefficient is determined by
\[
p_j(v)=-\frac{r_j(v)}{2(j+1-\rho)}.
\]
The denominator is nonzero because $\rho$ is a half-integer.
Continuing through $j=m-2$, we obtain
\[
\mathcal R[P]=s^{m-1}R(s,v),
\]
with $R$ analytic. Having constructed $P$, we introduce
the remainder $Y$ by
\[
v'=P(s,v)+s^{m-2}Y.
\]
This brings~\eqref{eq:endpoint-ode} into BFM form:
\begin{equation}
\begin{split}
sv'&=s\bigl(P(s,v)+s^{m-2}Y\bigr),\\
sY'&=-\frac12Y+sG(s,v,Y),
\end{split}
\label{eq:endpoint-analytic-BFM}
\end{equation}
where
\[
G(s,v,Y)=
\left(\frac{\rho-1}{2+s}-P_v(s,v)\right)Y
-\frac{R(s,v)}{2+s}
\]
is analytic. Here the coefficient $-\tfrac12$ follows from
$\rho-1-(m-2)=-\tfrac12$.
Theorem~\ref{thm:BFM} gives a solution with
$v(0)=c$ and $Y(0)=0$, analytic in $s$ and $c$.
We denote the resulting analytic solution of
Eq.~\eqref{eq:endpoint-ode} by $v_0(s;c)$.
Since the Taylor recurrence for~\eqref{eq:endpoint-ode} has
nonzero divisors $2n(n-\rho)$ for every integer $n\ge1$,
this is the unique analytic solution with endpoint value $c$.

\paragraph{The fractional-power family.}
To include the second free parameter, we write
\begin{equation}
v(s;c,B)=v_0(s;c)+s^\rho Z(\sigma;c,B),
\qquad \sigma=\sqrt{s}.
\label{eq:endpoint-family}
\end{equation}
In terms of $\sigma$, the factor $s^\rho$ becomes the integer
power $\sigma^{2m-3}$. Substituting into~\eqref{eq:endpoint-ode}
and writing $W=dZ/d\sigma$, we obtain
\begin{equation}
\begin{split}
\sigma\,\frac{dZ}{d\sigma}&=\sigma W,\\
\sigma\,\frac{dW}{d\sigma}
&=-(2\rho+1)W+\sigma H(\sigma,Z,W;c),
\end{split}
\label{eq:endpoint-BFM}
\end{equation}
where,
\[
\begin{split}
(2+\sigma^2)H={}&2(\rho-1)\sigma W
+4\bigl[\rho(\rho-1)-2(m-1)(1-3v_0^2)\bigr]Z\\
&+24(m-1)v_0\sigma^{2\rho}Z^2
+8(m-1)\sigma^{4\rho}Z^3.
\end{split}
\]
Since $2\rho=2m-3$ is an integer, $H$ is analytic.
The BFM theorem gives a unique solution with
\[
Z(0;c,B)=B,\qquad W(0;c,B)=0,
\]
analytic in $(\sigma,c,B)$. 
Thus $c$ fixes the endpoint value, while $B$ fixes the leading
fractional-power correction. For $B=0$, uniqueness gives $Z=W=0$,
and the solution reduces to $v_0(s;c)$.

To recover the decomposition used in the main text, split $Z$ into
its even and odd parts:
\[
Z(\sigma;c,B)
=
Z_{\mathrm e}(\sigma^2;c,B)
+\sigma Z_{\mathrm o}(\sigma^2;c,B).
\]
Since $\rho+\tfrac12=m-1$ is an integer, the odd part contributes
only integer powers of $s$. Moreover, $Z_{\mathrm e}$ vanishes
at $B=0$, so it has an analytic factor $B$. Consequently, the
endpoint family can be written as
\begin{equation}
v(s;c,B)
=v_{\mathrm{an}}(s;c,B)+Bs^\rho G_0(s;c,B),
\qquad G_0(0;c,B)=1,
\label{eq:endpoint-analytic-decomposition}
\end{equation}
where both functions $v_{\mathrm{an}}$ and $G_0$ are analytic
in $s$ and the parameters.
The analytic component need not be a solution, but equals $v_0$
when $B=0$.

Returning to $y$, we obtain
\[
u(y)=u_{\mathrm{an}}(y;c,B)
+B(1-y)^\rho g(y;c,B),
\]
where
\[
u_{\mathrm{an}}(1;c,B)=c,\qquad g(1;c,B)=1.
\]

\subsection{Identification of origin-regular shots}

We show that every origin-regular shot belongs to the family
constructed above. Fix one such shot, suppress its parameter $a$,
and let $c$ be its endpoint value, which exists by~\cite{BGW}.
Subtracting the equation for $v_0(s;c)$ from that for $v$ shows
that $\eta=v-v_0(s;c)$ satisfies the linear equation
\[
s(2+s)\eta''-2(\rho-1)(1+s)\eta'+Q(s)\eta=0,
\qquad
Q(s)=(2m-2)(1-v^2-vv_0-v_0^2),
\]
with $\eta(s)\to0$ as $s\to0^+$.

We seek a solution $\eta_1=s^\rho z$ with $z(0)=1$.
Substitution gives
\[
s(2+s)z''+2(\rho+1+s)z'
+\bigl[Q(s)-\rho(\rho-1)\bigr]z=0.
\]
An elementary local existence argument gives such a solution
with bounded $z'$, so $\eta_1(s)=s^\rho(1+O(s))$.
Reduction of order yields a second solution,
linearly independent of $\eta_1$,
\[
\eta_2(s)=\eta_1(s)\int^s
\frac{r^{\rho-1}(2+r)^{\rho-1}}{\eta_1(r)^2}\,dr
\sim-\frac{2^{\rho-1}}{\rho},
\qquad s\to0^+.
\]
Since $\eta\to0$, its $\eta_2$ component vanishes,
so $\eta=B\eta_1$ for a unique $B$.

Setting $\sigma=\sqrt{s}$, we therefore obtain
\[
Z(\sigma):=\sigma^{-2\rho}\eta(\sigma^2)
=Bz(\sigma^2)=B+O(\sigma^2),
\qquad
W(\sigma):=\frac{dZ}{d\sigma}
=2B\sigma z'(\sigma^2)=O(\sigma).
\]
These functions satisfy~\eqref{eq:endpoint-BFM}.
The uniqueness assertion in Theorem~\ref{thm:BFM} therefore
identifies the shot with the member of the family constructed
above corresponding to $(c,B)$.

\subsection{Dependence on the shooting parameter}

The endpoint value and fractional amplitude are genuine coordinates
for the local solution, not merely formal expansion coefficients.
To see this, fix a small $s_0>0$ and consider
\[
(c,B)\longmapsto
\bigl(v(s_0;c,B),\partial_sv(s_0;c,B)\bigr).
\]
The endpoint expansion gives
\[
\partial_cv=1+O(s),\qquad
\partial_s\partial_cv=O(1),
\]
and
\[
\partial_Bv=s^\rho(1+O(s)),\qquad
\partial_s\partial_Bv
=\rho s^{\rho-1}(1+O(s)).
\]
Consequently,
\[
\det\frac{\partial(v,v_s)}{\partial(c,B)}
=\rho s_0^{\rho-1}(1+O(s_0))\ne0.
\]
The analytic inverse function theorem shows that $(c,B)$ depend
locally real analytically on the interior value and derivative.
Since the origin-regular shot depends analytically on $a$ at
every interior point, we obtain the following result.

\begin{theorem}
\label{thm:endpoint-dependence}
The endpoint parameters $(c,B)$ depend locally real analytically
on the value and derivative of the solution at a fixed interior
point sufficiently close to $y=1$.
In particular, for the origin-regular shooting family,
$c(a)$ and $B_m(a)$ are locally real analytic functions of $a>0$.
An origin-regular shot is smooth at $y=1$ if and only if
$B_m(a)=0$. In that case, it is analytic there.
\end{theorem}

The smoothness criterion follows directly from the endpoint
expansion. When $B=0$, the solution coincides with the analytic
solution $v_0$. When $B\ne0$, the term
$B(1-y)^\rho g(y)$, with $g(1)=1$, prevents smoothness
because $\rho$ is a half-integer.

Near the zero solution, linearizing the map from interior data
to $B$ gives the fractional-power coefficient of the corresponding
linearized solution. This observation is used
in~\ref{app:matching} to transfer the outer approximation
to the endpoint amplitude.

\section{Matching the core to the outer solution}
\label{app:matching}\label{sec:matching}

For $m\ge6$, we justify the large-$a$ approximation
\[
u(y;a)\sim C_ma^{-k/2}F_m(y)
\]
and its consequence for the fractional-power coefficient,
\[
B_m(a)\sim 2^\rho C_m\mathcal K_m a^{-k/2}.
\]
The main point is that the nonlinear core fixes the amplitude
of the outer solution, while nonlinear effects outside the core
produce only a smaller correction.

\paragraph{Matching in the overlap region.}
Choose a large number $R$ and match at
\[
y_R=\frac{R}{\sqrt a}.
\]
For large $a$, this point lies well inside the light cone but
outside the core. Convergence to the core profile gives
\[
u(y_R;a)\simeq Q(R),
\qquad
y_Ru_y(y_R;a)\simeq RQ'(R).
\]
Let
\[
L_a(y)=\alpha_{a,R}F_m(y)+\beta_{a,R}G_m(y)
\]
be the solution of the linear outer equation with these same
value and derivative at $y_R$. Here
\[
F_m(y)\sim y^{-k},
\qquad
G_m(y)\sim y^{-\ell},
\qquad
\ell=m-2+\nu>k,
\]
as $y\to0$.
Matching the two powers gives, for fixed $R$ as $a\to\infty$,
\[
a^{k/2}\alpha_{a,R}
\longrightarrow
C_{m,R}:=
R^k\frac{\ell Q(R)+RQ'(R)}{\ell-k},
\qquad
a^{k/2}\beta_{a,R}\longrightarrow0.
\]
The core tail $Q(R)\sim C_mR^{-k}$ implies
$C_{m,R}\to C_m$ as $R\to\infty$.
Thus the core selects the outer mode $F_m$ with amplitude
$C_ma^{-k/2}$; the contribution of the faster mode $G_m$
is negligible at fixed $y>0$.

\paragraph{Size of the nonlinear correction.}
At the matching point, the field is already small:
$u(y_R;a)=O(R^{-k})$.
The cubic term is therefore smaller than the linear term
by a factor $O(R^{-2k})$.
To check that this small correction does not accumulate as
we move outwards, write the equation in integral form:
\[
u(y;a)-L_a(y)
=\int_{y_R}^y
\mathcal G_m(y,s)u(s;a)^3\,ds.
\]
For $y_R\le s\le y\le y_0$, where $y_0>0$ is fixed and small,
the linear Green function satisfies
\[
|\mathcal G_m(y,s)|\le Cy^{-k}s^{k-1}.
\]
Using the matched decay
$u(s;a)=O(a^{-k/2}s^{-k})$ in this integral gives
\[
|u(y;a)-L_a(y)|
\le
CR^{-2k}a^{-k/2}y^{-k}.
\]
For sufficiently large $R$, this estimate also makes the
assumed decay self-consistent. The derivative is controlled
in the same way at $y=y_0$.

Hence the nonlinear correction can be made arbitrarily small
relative to the leading outer field by choosing $R$ large.
Taking $a\to\infty$ first and then $R\to\infty$, we obtain
\[
a^{k/2}\bigl(u(y_0;a),u_y(y_0;a)\bigr)
\longrightarrow
C_m\bigl(F_m(y_0),F_m'(y_0)\bigr).
\]
Away from the origin and the light cone, ordinary continuous
dependence propagates this convergence. Thus
\[
a^{k/2}u(\,\cdot\,;a)\longrightarrow C_mF_m
\]
together with the first derivative on every closed subinterval
of $(0,1)$.

\paragraph{Transfer to the light cone.}
Convergence at interior points does not by itself determine
the fractional-power coefficient at the singular endpoint.
For this last step, we use the endpoint coordinates constructed
in~\ref{app:endpoint}.

Fix $y_1<1$ sufficiently close to the light cone.
Near the zero solution, the fractional-power coefficient is
an analytic function $\mathcal B$ of the interior data
$(u(y_1),u_y(y_1))$.
Its linear term gives the fractional-power coefficient of the
corresponding solution of the linearized equation.
Since
\[
\bigl(u(y_1;a),u_y(y_1;a)\bigr)
=
C_ma^{-k/2}
\bigl(F_m(y_1),F_m'(y_1)\bigr)
+o(a^{-k/2}),
\]
we need only the fractional-power coefficient of $F_m$.
By~\eqref{eq:Fconnection}, this is $2^\rho\mathcal K_m$.
Consequently,
\[
B_m(a)
=
2^\rho C_m\mathcal K_m a^{-k/2}
+o(a^{-k/2}),
\]
which proves~\eqref{eq:B-asymptotic}.

\section{Taylor coefficient estimates}
\label{app:coefficients}

We establish the eventual coefficient signs and the divergence of
$f^{(m-1)}$ used in Section~3. Throughout, $\rho=m-\frac32$,
and $A$ denotes the parameter of the explicit branch under
consideration.

\subsection{The recurrence and a comparison identity}

Substituting the Taylor series
\[
f(\xi)=\sum_{n\ge0}f_n\xi^n
\]
into the Heun equation~\eqref{heun} gives the recurrence
\begin{equation}
D_nf_{n+1}=T_nf_n-C_nf_{n-1},
\qquad n\ge1,
\label{eq:C-recurrence}
\end{equation}
where
\begin{equation}
\begin{split}
D_n&=A(n+1)(n+m+1),\\
T_n&=(A+1)n^2+
\left(m+\frac{5A+1}{2}\right)n
-m-\frac{3A-1}{2},\\
C_n&=(n+3)(n-2).
\end{split}
\label{eq:C-recurrence-coefficients}
\end{equation}
The starting coefficients are
\[
f_0=1,
\qquad
f_1=-\frac{3A+2m-1}{2A(m+1)},
\]
and
\[
f_2=
\frac{(A-1)(2m-1-3A)}
{2A^2(m+1)(m+2)},
\qquad
f_3=\frac{15A+2m+11}{6A(m+3)}f_2.
\]

Let $g_n$ denote either $f_n$ or $-f_n$. It satisfies the same
recurrence. To obtain lower bounds on $g_n/g_{n-1}$, choose a
positive sequence $r_n$ and set
\begin{equation}
e_n=g_n-r_ng_{n-1},
\qquad
E_n=T_n-\frac{C_n}{r_n}-D_nr_{n+1}.
\label{eq:C-comparison-definitions}
\end{equation}
Then \eqref{eq:C-recurrence} gives the exact identity
\begin{equation}
D_ne_{n+1}=\frac{C_n}{r_n}e_n+E_ng_n.
\label{eq:C-comparison-identity}
\end{equation}

Suppose that for some $N\ge3$,
\[
g_{N-1}>0,\qquad e_N>0,\qquad E_n>0\quad(n\ge N).
\]
Then $g_N=r_Ng_{N-1}+e_N>0$, and
\eqref{eq:C-comparison-identity} gives $e_{N+1}>0$,
since $C_n,D_n>0$ for $n\ge3$.
Induction therefore yields
\begin{equation}
g_n>r_ng_{n-1}>0
\qquad(n\ge N).
\label{eq:C-propagated-comparison}
\end{equation}
Iterating this inequality gives
\[
g_n>g_{N-1}\prod_{j=N}^n r_j.
\]
If
\[
r_n=1-\frac{\alpha}{n}+O(n^{-2}),
\]
then taking logarithms and summing gives
\[
\log\prod_{j=N}^n r_j=-\alpha\log n+O(1).
\]
Consequently,
\[
g_n > c_m n^{-\alpha}
\qquad(n\ge N)
\]
for some $c_m>0$. Thus it suffices to verify the starting
inequalities $g_{N-1}>0$ and $e_N>0$, together with
$E_n>0$ for $n\ge N$. Here $c_m>0$ is independent of $n$.

We treat the branches separately below. In each case, the
comparison identity determines the eventual coefficient sign.
For the negative tails, the resulting polynomial lower bound
also proves endpoint divergence. For the positive tail, we
first exclude analyticity at $\xi=1$ by comparing coefficient
ratios, and then deduce divergence from the endpoint expansion.

\subsection{Negative coefficient tails}

\paragraph{The plus branch for $m\ge5$.}

Set $g_n=-f_n$ and choose
\[
r_n=\frac{n-2}{n+m-\frac52},
\qquad n\ge3.
\]
On this branch $A>2m-3$, so $g_2,g_3>0$. Moreover,
\[
\frac{g_3}{g_2}-r_3
=
\frac{18Am-21A+4m^2+24m+11}
{6A(m+3)(2m+1)}
>0.
\]
Thus the comparison starts at $n=3$.

A direct calculation gives
\[
E_n=\frac{S_mn+Q_m}{2(2n+2m-3)},
\]
where
\[
S_m=A(10m-17)-16m+32,
\qquad
Q_m=A(13-2m)-16m^2+56m-48.
\]
Using $A>2m-3$, we find
\[
S_m>20m^2-80m+83>0,
\qquad
3S_m+Q_m>40m^2-152m+162>0.
\]
Hence
\[
S_mn+Q_m=(n-3)S_m+(3S_m+Q_m)>0
\qquad(n\ge3),
\]
so $E_n>0$ for all $n\ge3$.
The comparison identity therefore gives
\[
g_n>r_ng_{n-1}>0,\qquad n\ge3.
\]
Since
\[
r_n=1-\frac{\rho+1}{n}+O(n^{-2}),
\]
it follows that
\begin{equation}
-f_n\ge c_m n^{-\rho-1}
\label{eq:C-negative-tail-bound}
\end{equation}
for some $c_m>0$ and all $n\ge3$.

\paragraph{The minus branch for $m=6,7,8$.}

Set $g_n=-f_n$ and choose
\[
r_n=\frac{n-16}{n-16+m-\frac12},
\qquad n\ge66.
\]
The comparison starts later because $f_2,f_3>0$.
For each $m=6,7,8$, we use exact arithmetic in Maple to verify
the AI-suggested starting inequalities
\begin{equation}
g_{65}>0,\qquad
\frac{g_{66}}{g_{65}}-r_{66}>\frac1{1000},
\label{eq:C-finite-checks}
\end{equation}
so $e_{66}>0$. The same calculation shows that
\[
2(n-16)(2n+2m-31)E_n
\]
is a polynomial in $n-66$ with strictly positive coefficients.
Hence $E_n>0$ for all
$n\ge66$, and the comparison identity gives
\[
g_n>r_ng_{n-1}>0,\qquad n\ge66.
\]
Since
\[
r_n=1-\frac{\rho+1}{n}+O(n^{-2}),
\]
we again obtain \eqref{eq:C-negative-tail-bound}.

\paragraph{Nonvanishing of $\mathcal C$ and endpoint divergence.}

In all the negative-tail cases, the eventual coefficient sign
is $\sigma=-1$. Moreover,
\[
\frac{n!}{(n-m+1)!}\sim n^{m-1},
\qquad
m-2-\rho=-\frac12.
\]
It follows from \eqref{eq:C-negative-tail-bound} that
\[
\frac{n!}{(n-m+1)!}f_n
\le-\widetilde c_m n^{-1/2}
\]
for all sufficiently large $n$, with $\widetilde c_m>0$.
Consequently, the tail of the differentiated Taylor series
\[
f^{(m-1)}(\xi)
=\sum_{n\ge m-1}
\frac{n!}{(n-m+1)!}f_n\xi^{n-m+1}
\]
tends to $-\infty$ as $\xi\to1^-$.
The finitely many preceding terms remain bounded, so
\[
f^{(m-1)}(\xi)\longrightarrow-\infty
\qquad\mbox{as}\qquad \xi\to1^-.
\]
The endpoint expansion~\eqref{eq:f-endpoint-derivative}
therefore implies $\mathcal C\ne0$.

\subsection{The minus branch for $m\ge9$.}

Set $g_n=f_n$, write
\[
R_m=2\sqrt{3(m-2)(m-1)},
\qquad
A=\frac{4m-5-R_m}{3},
\]
and choose
\begin{equation}
r_n=
\frac{n+1}{A(n+\frac52)}
\left(1+\frac1{3n^2}\right),
\qquad n\ge3.
\label{eq:C-positive-comparison}
\end{equation}
We first prove positivity of the coefficients and then establish
nonvanishing of $\mathcal C$ and endpoint divergence.

\paragraph{Starting the comparison.}

Here $g_2>0$, and
\[
\frac{g_3}{g_2}-r_3
=
\frac{5(297A-50m-51)}{594A(m+3)}.
\]
Substituting the formula for $A$ gives
$297A-50m-51>0$ for $m\ge9$.
Thus $g_3/g_2>r_3$, initializing the comparison at $n=3$.

\paragraph{Positivity of the gap.}

We first establish the lower bound
\[
A>L(m):=\frac{22}{13}+\frac4{23}(m-9),
\qquad m\ge9.
\]
Indeed, direct calculation gives
$A(9)=(31-4\sqrt{42})/3>22/13$ and
\[
A'(m)=\frac23\left(2-\frac{6m-9}{R_m}\right)>\frac4{23}.
\]
Integration yields the claim.

For \eqref{eq:C-positive-comparison}, direct calculation gives
\[
E_n=
\frac{P_m(n;A)}
{6(n+1)(2n+7)(3n^2+1)},
\]
where
\[
\begin{split}
P_m(n;A)={}&(174A-18m-111)n^4
+(1137A-228m-144)n^3\\
&+(1860A-228m-70)n^2
+(24A-76m-48)n-63A-74m-11.
\end{split}
\]
Since
\[
\partial_A P_m(n;A)
=3(n+3)(2n+7)(29n^2+n-1)>0
\qquad(n\ge3),
\]
it suffices to substitute $A=L(m)$. At this value,
\[
P_m(n;L(m))
=\frac{J(n)}{13}
+\frac{2(m-9)}{23}K(n-3),
\]
where
\[
J(n)=279n^4-3534n^3+13334n^2-8988n-10187
\]
and
\[
K(j)=141j^4+1344j^3+5580j^2+11594j+8452.
\]
All coefficients of $K$ are positive, as are
$J(3)=10036$, $J(4)=12453$, $J(5)=10848$, and
\[
J(6+j)=279j^4+3162j^3+9986j^2+10404j+14149
\qquad(j\ge0).
\]
Hence $E_n>0$ for every integer $n\ge3$.
The comparison identity therefore gives
\[
g_n>r_ng_{n-1}>0,\qquad n\ge3.
\]
Since $g_n=f_n$, the eventual coefficient sign is $\sigma=1$.
Moreover, the comparison product is asymptotic to a positive
constant times $A^{-n}n^{-3/2}$, so
\[
g_n > c_m A^{-n}n^{-3/2}
\qquad n\ge3,
\]
where $c_m>0$ is independent of $n$.

\paragraph{Nonvanishing of $\mathcal C$ and endpoint divergence.}

Returning to $f_n=g_n$, we have
\[
f_n > c_m A^{-n}n^{-3/2}.
\]
This bound does not by itself force divergence at $\xi=1$.

We prove that $\mathcal C\ne0$ by contradiction.
Suppose that $\mathcal C=0$. Then $f$ is analytic at $\xi=1$,
so its Taylor series at $0$ has radius at least $A$.
The lower bound above shows that the radius is exactly $A$.
Thus $f$ is singular at $\xi=A$, where the local exponents
are $0$ and $\frac12$, and hence has a nonzero square-root
component there. Standard coefficient asymptotics give
a ratio expansion whose $n^{-2}$ coefficient is determined
by the recurrence:
\[
\frac{f_n}{f_{n-1}}
=\frac1A\left(1-\frac{3}{2n}
+\frac{2m-4-3A}{4(A-1)n^2}+O(n^{-3})\right).
\]
On the other hand,
\[
r_n=\frac1A\left(1-\frac{3}{2n}
+\frac{49}{12n^2}+O(n^{-3})\right).
\]
The $n^{-2}$ coefficient in the ratio expansion is strictly
smaller than that in $r_n$, since
\[
\frac{49}{12}-\frac{2m-4-3A}{4(A-1)}
=\frac{58A-6m-37}{12(A-1)}>0,
\]
where the last inequality follows from $A>L(m)$.
Thus $f_n/f_{n-1}<r_n$ for all sufficiently large $n$,
contrary to the comparison inequality.
Hence $\mathcal C\ne0$.

By~\eqref{eq:f-endpoint-derivative} and eventual positivity
of the Taylor coefficients,
\[
f^{(m-1)}(\xi)\longrightarrow+\infty
\qquad\mbox{as}\qquad \xi\to1^-.
\]
This completes the coefficient and divergence estimates
used in Section~3.

\section*{Acknowledgements}

AI assistance was used in Appendix~C to explore recurrence relations,
choose comparison ratios, initialize the induction arguments,
and perform exact-arithmetic calculations. The author takes
full responsibility for the mathematical content and final text.


\begin{thebibliography}{9}
\bibitem{BGW}
P.~Bizo\'n, I.~Glogi\'c, and A.~Wasserman,
\emph{Three self-similar solutions of Yang--Mills equations in high odd dimensions}, Phys.\ Rev.\ D \textbf{114}, 064021 (2026).
\bibitem{BFM}
P.~Breitenlohner, P.~Forg\'acs, and D.~Maison,
\emph{Static Spherically Symmetric Solutions of the Einstein--Yang--Mills Equations}, Commun.\ Math.\ Phys.\ \textbf{163} (1994), 141--172.
\bibitem{BK}
P.~Bizo\'n and R.~Kluszczy\'nski,
\emph{Existence of a third self-similar solution of the Yang--Mills equation in high odd dimensions},
arXiv:2610.36784 (2026).
\bibitem{DLMF}
NIST Digital Library of Mathematical Functions,
\url{https://dlmf.nist.gov/}, Eq.~(15.8.4).
\end{thebibliography}
\end{document}